\documentclass[sigconf]{acmart}

\usepackage{graphicx}
\usepackage{caption}
\AtBeginDocument{%
  }

\copyrightyear{2026}
\acmYear{2026}
\setcopyright{cc}
\setcctype{by-nc-nd}
\acmConference[CHI EA '26]{Extended Abstracts of the 2026 CHI Conference on Human Factors in Computing Systems}{April 13--17, 2026}{Barcelona, Spain}
\acmBooktitle{Extended Abstracts of the 2026 CHI Conference on Human Factors in Computing Systems (CHI EA '26), April 13--17, 2026, Barcelona, Spain}
\acmDOI{10.1145/3772363.3798953}
\acmISBN{979-8-4007-2281-3/2026/04}

\acmArticleType{Research}
\begin{document}

\title{``To LLM, or Not to LLM'': How Designers and Developers Navigate LLMs as Tools or Teammates}

\author{Varad Vishwarupe}
\email{varad.vishwarupe@cs.ox.ac.uk}
\affiliation{%
  \institution{Department of Computer Science, University of Oxford}
  \city{Oxford}
  \country{United Kingdom}
}

\author{Ivan Flechais}
\email{ivan.flechais@cs.ox.ac.uk}
\affiliation{%
  \institution{Department of Computer Science, University of Oxford}
  \city{Oxford}
  \country{United Kingdom}
}

\author{Nigel Shadbolt}
\email{nigel.shadbolt@cs.ox.ac.uk}
\affiliation{%
  \institution{Department of Computer Science, University of Oxford}
  \city{Oxford}
  \country{United Kingdom}
}

\author{Marina Jirotka}
\email{marina.jirotka@cs.ox.ac.uk}
\affiliation{%
  \institution{Department of Computer Science, University of Oxford}
  \city{Oxford}
  \country{United Kingdom}
}


\renewcommand{\shortauthors}{Vishwarupe et al.}

\begin{abstract}
Large language models (LLMs) are increasingly integrated into design and development workflows, yet decisions about their use are rarely binary or purely technical. We report findings from a constructivist grounded theory study based on interviews with 33 designers and developers across three large technology organisations. Rather than evaluating LLMs solely by capability, participants reasoned about the role an LLM could occupy within a workflow and how that role would interact with existing structures of responsibility and organisational accountability. When LLMs were framed as tools under clear human control, their use was typically acceptable and could be integrated within existing governance structures. When framed as teammates with shared or ambiguous agency, practitioners expressed hesitation, particularly when responsibility for outcomes could not be clearly justified. At the same time, participants also described productive teammate configurations in which LLMs supported collaborative reasoning while remaining embedded within explicit oversight structures. We identify tool and teammate framings as recurring ways in which designers and developers position LLMs relative to human work and present an analytic rubric describing how role framing shapes decision authority, accountability ownership, oversight strategies, and organisational acceptability. By foregrounding design-time reasoning, this work reframes ``To LLM or Not to LLM'' as a sociotechnical positioning problem that emerges during system design rather than during post-deployment evaluation.
\end{abstract}



\maketitle

\section{Introduction}
Large language models (LLMs) are increasingly embedded in contemporary product design and software development workflows. They now support activities ranging from code generation and content authoring to summarisation and decision support. As these models become readily available, designers and developers are increasingly required to decide not only whether to use an LLM, but how it should be situated within human work. In practice, integrating an LLM requires negotiating responsibility, accountability, and acceptable forms of control within organisational settings where humans remain responsible for outcomes. Designers and developers must consider how model outputs will be interpreted, reviewed, and justified in contexts shaped by product risk, governance processes, and professional accountability. As a result, decisions about LLM use often involve uncertainty and negotiation rather than simple assessments of accuracy or performance.

Human-computer interaction research has long examined how people interact with intelligent systems, how trust and reliance develop, and how mental models shape collaborative activity with automation and AI. This body of work has produced important insights into how system behaviour, transparency, and reliability influence interaction and reliance~\cite{lee2004trust,hoff2015trust}. However, much of this research focuses on use-time interaction or post-adoption evaluation, leaving the design-time reasoning of practitioners comparatively underexplored. Comparatively less attention has been given to design-time reasoning, where practitioners must determine how an AI system should be positioned relative to human work before it is deployed. In these contexts, practitioners negotiate how authority, accountability, and oversight will be distributed between humans and AI systems within organisational workflows.

In this paper, we focus on designers' and developers' reasoning about LLM integration. Specifically, we examine how practitioners reason about the roles LLMs might play in relation to human work, and how these role framings shape decisions to adopt, constrain, or abandon LLM use. To do so, we adopt a constructivist grounded theory approach that allows analytic categories to emerge inductively from practitioners' accounts rather than being imposed in advance~\cite{charmaz2014constructing}. We report findings from interviews with 33 designers and developers across three large technology organisations. Our analysis shows that participants reasoned about LLM integration by considering whether an LLM could be positioned as a tool under clear human control or as a teammate with shared or ambiguous agency. We then report on follow-up interviews that examine how this distinction shapes accountability ownership, oversight strategies, and organisational acceptability, resulting in a rubric that characterises the practical implications of different role framings. Together, this work reframes ``To LLM or Not to LLM'' as a positioning problem rather than a purely technical decision, contributing an empirically grounded account of how LLM roles are negotiated during design and development practices.

\section{Literature Review and Motivation}

HCI research has increasingly moved beyond treating AI systems as passive automation, instead examining how they participate in human work. A growing literature on human-AI teaming explores how intelligent systems can support joint activity, coordinate with humans, and contribute to shared goals. This work highlights challenges related to coordination, role clarity, and the distribution of initiative, often framing AI systems as collaborators or teammates rather than simple tools~\cite{shneiderman2020human,amershi2019guidelines}. Alongside this, HCI researchers have extensively studied trust, reliance, and calibration in human-automation and human-AI interaction. Foundational work on trust in automation demonstrates that appropriate reliance depends not only on system performance, but also on users' understanding of system capabilities and the clarity of responsibility when failures occur~\cite{lee2004trust}. Subsequent research shows that mismatches between system behaviour and user understanding can lead to over-reliance, under-reliance, or brittle interaction patterns~\cite{hoff2015trust,suchman2007human}.

More recent studies extend these concerns to AI and LLM-based systems, showing that fluent, human-like outputs can complicate judgments about reliability and appropriate oversight~\cite{amershi2019guidelines}. Research on mental models further demonstrates that how people conceptualise intelligent systems shapes how they interact with them. Users' interpretations of system agency and competence influence reliance, error attribution, and responsibility assignment~\cite{liao2020questioning,eiband2018bringing,vishwarupe2022xai,odriscoll2025socialnorms,vishwarupe2016webbehavior}. In the context of LLMs, emerging evidence suggests that variability in outputs and limited transparency make it difficult for people to form stable mental models, complicating decisions about appropriate use and oversight~\cite{kaur2020interpreting,bansal2021whole,vishwarupe2022hcai,vishwarupe2018twitterspam}.

At the same time, dominant models of technology adoption and continuance in information systems research, such as Expectation Confirmation Theory and the Expectation-Confirmation Model, primarily explain post-adoption satisfaction and continued use in terms of perceived usefulness and confirmation of expectations~\cite{oliver1980cognitive,bhattacherjee2001understanding}. While valuable for understanding technology acceptance after deployment, these models are less well suited to capturing how designers and developers reason before adoption, particularly when AI systems can be integrated in multiple ways and when accountability structures are still being negotiated. Taken together, existing research provides strong accounts of interaction, trust, and collaboration with AI systems, but offers limited empirical insight into how practitioners decide what role an AI system should play during design and development. In practice, designers and developers must reason not only about what an LLM can do, but about how it should relate to human judgment, responsibility, and organisational accountability~\cite{passi2019problem,madaio2020codesigning,vishwarupe2023industry4,sayyed2025blocksafe,vishwarupe2025mentalhealth}.

In our grounded analysis of interview data, we observed a recurring distinction in how designers and developers reasoned about LLM integration. Rather than approaching LLM use as a simple yes-or-no decision, participants first considered whether an LLM could be positioned as a tool under clear human control or as a teammate with shared or ambiguous agency. This distinction emerged through iterative coding and constant comparison rather than being specified in advance. Given this recurring distinction, in this paper we ask the following Research Question:\\
\noindent\textbf{RQ:} How do designers and developers reason about whether and how to position LLMs as tools or teammates in practice?

\section{Methods}
We adopt a constructivist grounded theory (CGT) approach~\cite{charmaz2014constructing} because it allows analytic categories to emerge from participants' accounts rather than from prior frameworks or hypotheses. Our goal is not to evaluate LLM performance or predict usage outcomes but to generate a substantive theory that explains how practitioners navigate LLM adoption through role positioning. Interview prompts and illustrative coding examples are provided in Appendices A and B (supplementary materials).

\subsection{Study design and participants}
We conducted 33 semi-structured interviews with practitioners who design, develop or deploy AI-enabled systems. Participants included product designers, UX designers and researchers, software engineers and applied AI engineers. Professional experience ranged from 2 to more than 15 years, spanning early-career, senior and staff-level roles. Recruitment intentionally covered both design- and engineering-oriented positions to capture variation in decision authority, technical responsibility and organisational accountability. Participants were drawn from three globally leading multinational technology organisations with mature AI product portfolios. To preserve anonymity, we describe these organisations only at a high level as: a large consumer technology company, a cloud and enterprise software provider, and a search and AI infrastructure company. Sampling followed CGT principles. Data collection and analysis proceeded in parallel; emerging insights guided subsequent recruitment and questioning. We did not confine recruitment to a single product area or organisational unit because our interest lay in cross-cutting patterns in how LLM roles were reasoned about across different domains and teams. Recruitment continued until theoretical saturation---additional interviews no longer produced substantively new insights relevant to the developing theory.

\subsection{Data collection}
Interviews lasted 55--74 minutes and were conducted in-person or remotely. The protocol was intentionally open-ended and exploratory, focusing on participants' experiences with LLMs, moments of hesitation or uncertainty, and how decisions about LLM integration were discussed, justified or constrained within teams. Crucially, we did not introduce or presuppose role distinctions such as tool or teammate. Participants were invited to describe concrete situations in which LLMs were considered, used, modified, constrained or set aside and to explain how responsibility, oversight and justification were handled in those situations. This design choice reflects CGT's emphasis on allowing analytic categories to emerge inductively from participants' accounts rather than being imposed \emph{a priori}. Interviews were audio-recorded with consent, transcribed verbatim and anonymised. Participants are referenced using identifiers such as P7 (Software Engineer) or P14 (UX Researcher). This study received approval from the authors' institutional research ethics committee after due diligence. All participants provided informed written and verbal consent, and all data were anonymised prior to analysis. This study was conducted in accordance with institutional and ACM guidelines for research involving human participants and no personally identifiable or sensitive data were collected.

\subsection{Data analysis}
Analysis followed the core practices of constructivist grounded theory: iterative coding, memo writing and reflexive analysis. We began with open coding, attending closely to participants' language and reasoning when describing LLM use, hesitation, redesign or rejection. Initial codes captured how participants discussed control, responsibility, delegation, justification and organisational acceptability. Through constant comparison across interviews we compared instances of LLM use with instances of hesitation, redesign or non-use, examining how participants differentiated between acceptable and unacceptable configurations. Analysis focused on recurring concerns and decision rationales rather than on classifying participants or systems. We then engaged in focused coding to refine emerging analytic patterns, examining how participants reasoned about role boundaries, responsibility allocation and oversight practices across different contexts. Analytic memos helped explore relationships between codes, surface tensions and interrogate alternative interpretations.

To increase transparency regarding the analytic process, coding and memo writing were conducted by two researchers trained in qualitative methods. The first author conducted initial open coding across all transcripts to capture participants' language and decision reasoning. The second researcher independently reviewed a subset of transcripts and coding decisions to challenge emerging interpretations and support analytic consistency. Disagreements were resolved through iterative discussion and memo writing. This collaborative process helped refine analytic categories while maintaining sensitivity to participants' accounts.

Participants referenced a range of commercially available and internal LLMs, but our analysis intentionally abstracts away from specific models to focus on how practitioners reason about roles, accountability, and control across contexts.We report findings at the level of role framings and accountability configurations rather than at the level of particular model families or vendors. This abstraction reflects our analytic focus on design-time reasoning and organisational justifiability, and reduces the risk of over-interpreting model-specific behaviour that may change rapidly over time.

\subsection{Analytic refinement and follow-up interviews}
As the analytic categories took shape, we sought to refine and clarify them by conducting follow-up interviews with a subset of 15 participants from the original sample. Consistent with CGT principles, their purpose was analytic refinement---to densify and specify the emerging theory. In these interviews, participants reviewed anonymised summaries of emerging analytic patterns and reflected on similar reasoning in their own work. Insights from these discussions were integrated through constant comparison, helping to sharpen the dimensions that characterise how practitioners reason about LLM roles in practice. These dimensions inform the analytic rubric presented later in Table~\ref{tab:rubric}.

\subsection{Reflexivity and positionality}
The research team's background in HCI and AI system design informed sensitivity to design practices, organisational constraints, and accountability concerns. To avoid projecting prior assumptions onto the data, we engaged in active reflexivity through memo writing, regular analytic discussions, and deliberate consideration of alternative interpretations during coding. Particular attention was paid to moments where researchers' expectations diverged from participants' accounts, using these tensions as analytic resources for the theory to remain grounded in participants' reasoning practices.

\section{Findings and Discussion}
\begin{figure}[tb]
\centering
\includegraphics[width=\linewidth]{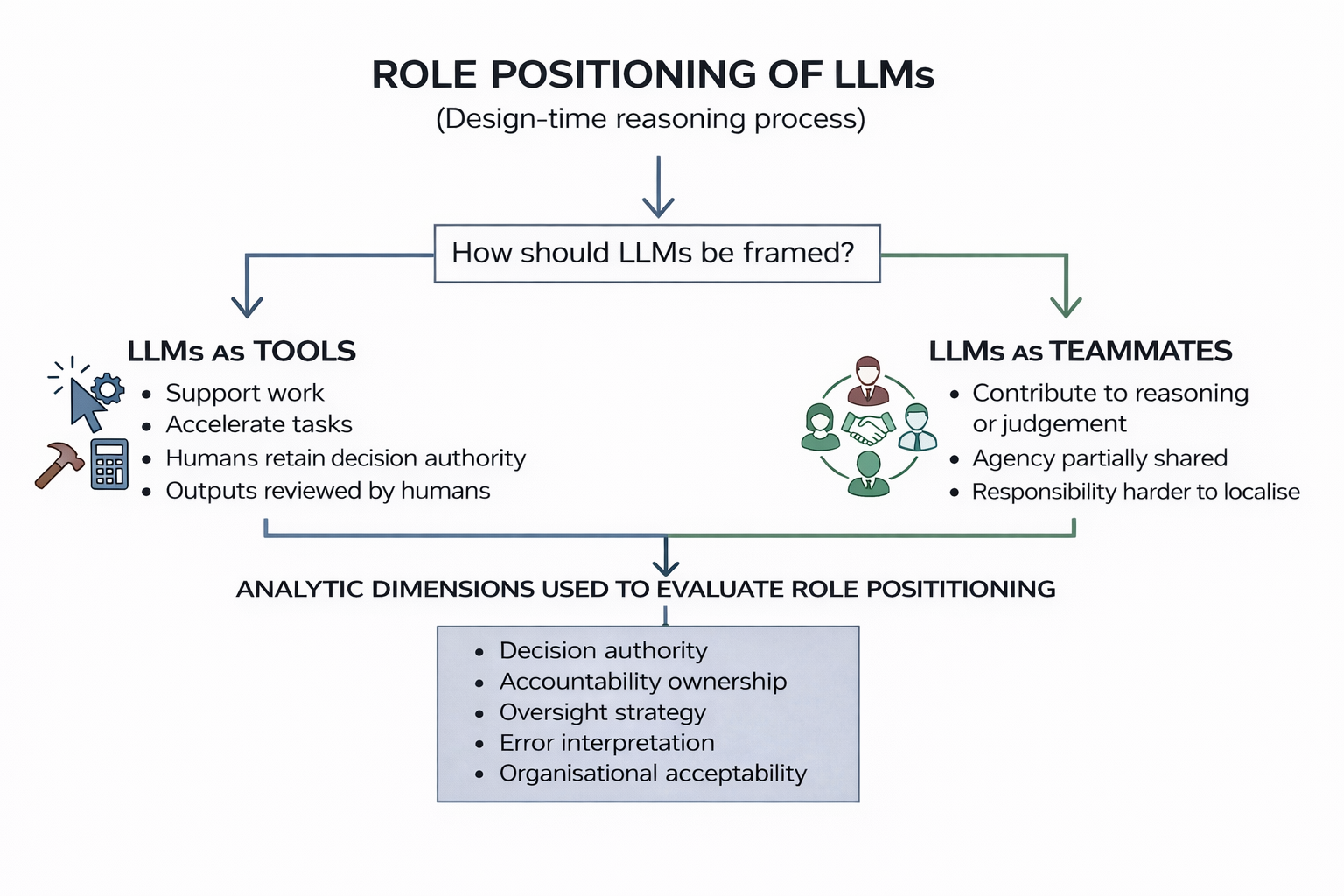}
\caption{Role positioning of LLMs as tools or teammates in design-time reasoning.}
\Description{Conceptual diagram illustrating how designers position LLMs along a continuum between tool-like systems under human control and teammate-like systems that contribute more autonomously to decision making.}
\label{fig:rolepositioning}
\end{figure}

Designers' and developers' decisions about whether to use an LLM are organised around a process of role positioning. Participants did not merely evaluate LLMs by capability; they reasoned about what role it could occupy within a workflow and whether that role could be reconciled with expectations of responsibility, accountability and organisational risk. Fig.~\ref{fig:rolepositioning} summarises this grounded process, showing how designers and developers navigate LLM adoption through role positioning. Additional examples of the interview protocol and coding process are provided in Appendices~A and~B.

\begin{figure}[t]
\centering
\includegraphics[width=\linewidth]{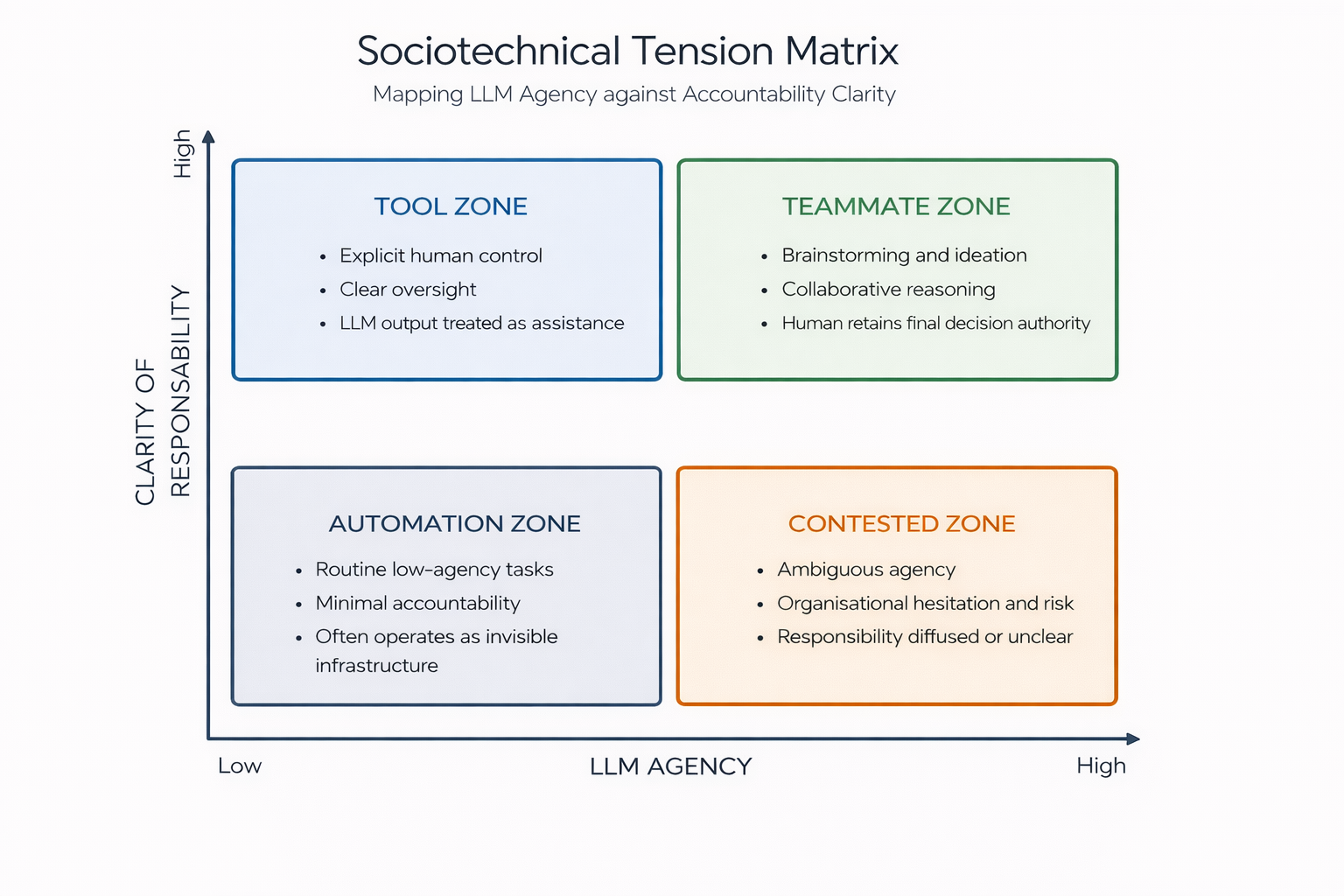}
\caption{Sociotechnical tension matrix mapping LLM agency against accountability clarity, highlighting tool, teammate, automation, and contested configurations.}
\Description{Two-dimensional matrix plotting levels of LLM agency against clarity of human accountability, identifying regions corresponding to tool use, teammate collaboration, automated delegation, and contested responsibility configurations.}
\label{fig:tensionmatrix}
\end{figure}

\begin{table}[t]
\centering
\caption{Analytic dimensions shaping role positioning in LLM integration.}
\label{tab:rubric}
\begin{tabular}{p{0.23\linewidth}p{0.34\linewidth}p{0.34\linewidth}}
\toprule
\textbf{Dimension} & \textbf{Tool framing} & \textbf{Teammate framing} \\
\midrule
Decision authority & Human-led; LLMs provide suggestions or drafts & Distributed or opaque; LLMs contribute to reasoning or judgments \\
Accountability ownership & Clearly human-owned and justifiable & Diffuse or contested; ownership difficult to localize \\
Oversight strategy & Explicit review, checkpoints and constraints & Difficult to formalize or enforce \\
Error interpretation & Treated as tool limitation or misuse & Responsibility for errors is ambiguous \\
Organisational acceptability & Often acceptable within existing governance & Frequently contested or delayed \\
\bottomrule
\end{tabular}
\end{table}

\subsection{Framing LLMs as tools: preserving control and accountability}
When participants described LLM use as acceptable, they most often framed the system as a tool---a resource that supports or accelerates human work while remaining subordinate to human judgment. In this framing, humans retain decision authority and are responsible for outcomes. LLM outputs are treated as provisional inputs to be reviewed, filtered or overridden. Oversight practices, such as human-in-the-loop review, constrained task scope and clear ownership of final decisions, were consistently associated with tool framing. Participants reported deliberately designing interfaces and workflows to reinforce this relationship. Errors were interpreted as limitations of a tool rather than failures of judgment, and organisational comfort with deployment was higher. For example, P7 (Software Engineer) said:

\begin{quote}
``If it's suggesting things and I'm still deciding, that feels fine. I'm accountable, so I need to be able to stand behind the decision.''
\end{quote}

\subsection{Framing LLMs as teammates: ambiguity, hesitation, and productive collaboration}
Hesitation often arose when LLMs were framed as teammates---when they were expected to contribute substantively to reasoning, interpretation or decision-making rather than merely supporting human work. Participants used collaborative language (``working with,'' ``pairing with'') to describe such configurations, but they also articulated discomfort around responsibility and accountability. Teammate framing introduced difficulty in justifying decisions, ambiguity about error ownership and heightened organisational or reputational risk. Participants described delaying, constraining or avoiding technically capable systems when they could not establish clear accountability structures, a tension summarised in Fig.~\ref{fig:tensionmatrix}. P14 (UX Researcher) remarked:

\begin{quote}
``Once it starts feeling like a collaborator, it's not obvious who owns the outcome anymore. That's where people get nervous.''
\end{quote}

At the same time, participants also described contexts in which teammate framings were productive, particularly for exploratory work such as ideation, early-stage design exploration, and brainstorming where outputs remain provisional and subject to human interpretation. In these cases, the LLM was valued as a cognitive partner that expanded the range of ideas considered, provided alternative framings, or accelerated iteration---so long as final judgment remained clearly human-owned. As one participant put it:

\begin{quote}
``It's useful when it feels like brainstorming with another engineer. But you still need to be clear that the final reasoning is yours.'' (P18, Software Engineer)
\end{quote}

These accounts suggest that teammate framings are not inherently undesirable but become organisationally difficult when collaborative contributions are interpreted as shared agency without clear oversight and responsibility.

\subsection{Navigating between roles}
Participants did not view tool and teammate framings as static or mutually exclusive categories. Instead, they described actively navigating between these positions, redesigning workflows, interfaces or governance processes to achieve a role alignment that felt acceptable. Strategies included narrowing task scope, adding additional review steps or restructuring workflows to ensure that humans remained visibly in control. Sometimes these efforts successfully shifted an LLM toward a more tool-like role; in other cases, participants concluded that reframing was not feasible. When role clarity could not be achieved, teams often chose not to proceed with LLM integration, despite recognising potential performance benefits. P28 (Applied AI Engineer) observed:

\begin{quote}
``The breakthrough wasn't making the model better. It was making the role clearer.''
\end{quote}

\subsection{Analytic dimensions of role positioning}
Follow-up interviews clarified how this role-positioning process operates in practice. While Fig.~\ref{fig:rolepositioning} captures the overall process, we found that participants consistently reasoned about LLM roles along a small set of recurring analytic dimensions. These dimensions include decision authority, accountability ownership, oversight strategies, error interpretation, and organisational acceptability. 
Table~\ref{tab:rubric} summarises these dimensions and illustrates how they vary depending on whether an LLM is framed as a tool or as a teammate. Rather than functioning as a prescriptive checklist, the rubric operates as an analytic device that captures the implicit reasoning practitioners use when negotiating LLM roles. Participants did not apply these dimensions as a formal evaluation instrument; instead, they surfaced repeatedly as teams debated whether a proposed LLM configuration could be reconciled with existing accountability structures. In several cases, teams redesigned systems or processes specifically to shift an LLM along these dimensions toward a more acceptable configuration. Together, these dimensions explain how decisions about LLM use are resolved in practice: by negotiating whether an acceptable alignment between human authority, system behaviour, and organisational responsibility can be established.

\section{Conclusion}
This paper contributes a constructivist grounded theory and a 5-dimensional analytic rubric explaining how designers and developers navigate decisions about LLM adoption through a process of role positioning rather than through binary or performance-driven judgments. We show that practitioners do not simply ask whether an LLM is capable, but whether it can be positioned in an acceptable role relative to human judgment, responsibility, and organisational accountability. When an LLM can be framed as a tool under clear human control, its use is often enabled; when it can only be framed as a teammate with shared or ambiguous agency, practitioners become cautious and may constrain or abandon its use altogether.

By foregrounding design-time reasoning, this work extends prior research on human-AI collaboration that has largely focused on interaction after deployment. Instead, we show that many consequential decisions about human-AI collaboration occur earlier, when practitioners determine what role an AI system should play within organisational workflows. Understanding these role negotiations is essential for designing AI systems that are not only technically capable but also accountable and organisationally acceptable.

\section{Limitations and Future Work}
This study draws on interviews with designers and developers from three large technology organisations and the resulting theory reflects practices and constraints characteristic of mature corporate settings; findings may not directly generalise to startups, public sector contexts, or non-Western organisational environments. As with all grounded theories, the account presented here is contextual and interpretive, capturing practitioners' reasoning at a particular moment in the evolution of LLM capabilities and governance frameworks. Future work could extend this theory across organisational scales, industries, and cultural settings to examine how role-positioning practices vary or evolve. Longitudinal studies could track how role configurations change over time as models, policies, and norms mature, while design-oriented research could investigate interventions---such as interface cues, workflow structures, or governance mechanisms---that support clearer role positioning and accountability in human-AI collaboration.

\bibliographystyle{ACM-Reference-Format}
\bibliography{REFERENCES}

@article{lee2004trust,
  author  = {John D. Lee and Katrina A. See},
  title   = {Trust in Automation: Designing for Appropriate Reliance},
  journal = {Human Factors},
  volume  = {46},
  number  = {1},
  year    = {2004},
  pages   = {50--80},
  doi     = {10.1518/hfes.46.1.50_30392}
}

@article{hoff2015trust,
  author  = {Kevin A. Hoff and Masooda Bashir},
  title   = {Trust in Automation: Integrating Empirical Evidence on Factors That Influence Trust},
  journal = {Human Factors},
  volume  = {57},
  number  = {3},
  year    = {2015},
  pages   = {407--434},
  doi     = {10.1177/0018720814554227}
}

@book{charmaz2014constructing,
  author    = {Kathy Charmaz},
  title     = {Constructing Grounded Theory},
  edition   = {2},
  publisher = {SAGE Publications Ltd},
  year      = {2014},
  address   = {London, United Kingdom}
}

@article{shneiderman2020human,
  author  = {Ben Shneiderman},
  title   = {Human-Centered Artificial Intelligence: Reliable, Safe \& Trustworthy},
  journal = {International Journal of Human-Computer Interaction},
  volume  = {36},
  number  = {6},
  year    = {2020},
  pages   = {495--504},
  doi     = {10.1080/10447318.2020.1741118}
}

@inproceedings{amershi2019guidelines,
  author    = {Saleema Amershi and Dan Weld and Mihaela Vorvoreanu and Adam Fourney and Besmira Nushi and Penny Collisson and Jina Suh and Shamsi Iqbal and Paul N. Bennett and Kori Inkpen and Thomas Teevan and Ruth Kikin-Gil and Eric Horvitz},
  title     = {Guidelines for Human-{AI} Interaction},
  booktitle = {Proceedings of the 2019 {CHI} Conference on Human Factors in Computing Systems},
  year      = {2019},
  publisher = {Association for Computing Machinery},
  address   = {New York, NY, USA},
  doi       = {10.1145/3290605.3300233},
  articleno = {3},
  numpages  = {13}
}

@book{suchman2007human,
  author    = {Lucy A. Suchman},
  title     = {Human-Machine Reconfigurations: Plans and Situated Actions},
  edition   = {2},
  publisher = {Cambridge University Press},
  address   = {Cambridge, United Kingdom},
  year      = {2007}
}

@inproceedings{liao2020questioning,
  author    = {Q. Vera Liao and Daniel Gruen and Sarah Miller},
  title     = {Questioning the {AI}: Informing Design Practices for Explainable {AI} User Experiences},
  booktitle = {Proceedings of the 2020 {CHI} Conference on Human Factors in Computing Systems},
  year      = {2020},
  publisher = {Association for Computing Machinery},
  address   = {New York, NY, USA},
  doi       = {10.1145/3313831.3376590},
  articleno = {416},
  numpages  = {15}
}

@inproceedings{eiband2018bringing,
  author    = {Malin Eiband and Daniel Buschek and Heinrich Hussmann and Alexander Butz},
  title     = {Bringing Transparency Design into Practice},
  booktitle = {Proceedings of the 23rd International Conference on Intelligent User Interfaces},
  year      = {2018},
  publisher = {Association for Computing Machinery},
  address   = {New York, NY, USA},
  doi       = {10.1145/3172944.3172961},
  pages     = {211--223}
}

@inproceedings{kaur2020interpreting,
  author    = {Himanshu Kaur and Harsha Nori and Samuel Jenkins and Rich Caruana and Hanna Wallach and Jennifer Wortman Vaughan},
  title     = {Interpreting Interpretability: Understanding Data Scientists' Use of Interpretability Tools for Machine Learning},
  booktitle = {Proceedings of the 2020 {CHI} Conference on Human Factors in Computing Systems},
  year      = {2020},
  publisher = {Association for Computing Machinery},
  address   = {New York, NY, USA},
  doi       = {10.1145/3313831.3376212},
  articleno = {280},
  numpages  = {14}
}

@incollection{sayyed2025blocksafe,
  author    = {Haazique Sayyed and Meshari Alwazae and Varad Vishwarupe},
  title     = {BlockSafe: Universal Blockchain-Based Identity Management},
  booktitle = {Big Data in Finance: Transforming the Financial Landscape: Volume 2},
  publisher = {Springer Nature Switzerland},
  address   = {Cham},
  year      = {2025},
  pages     = {57--66}
}

@incollection{vishwarupe2025mentalhealth,
  author    = {Varad Vishwarupe and Alexander Hankey and Shailesh Pangaonkar and Shwetanshu Shekhar and R. Sheena Rani and Meshari Alwazae and Haazique Sayyed and Vishal Pawar and Vidya Kamma and Priyanka Kuklani},
  title     = {Predicting Mental Health Ailments Using Social Media Activities and Keystroke Dynamics with Machine Learning},
  booktitle = {Big Data in Finance: Transforming the Financial Landscape: Volume 2},
  publisher = {Springer Nature Switzerland},
  address   = {Cham},
  year      = {2025},
  pages     = {33--44},
  doi       = {10.1007/978-3-031-80656-8_4}
}

@inproceedings{vishwarupe2018twitterspam,
  author    = {Varad Vishwarupe and Mangesh Bedekar and Milind Pande and Anil Hiwale},
  title     = {Intelligent Twitter Spam Detection: A Hybrid Approach},
  booktitle = {Smart Trends in Systems, Security and Sustainability},
  publisher = {Springer Singapore},
  address   = {Singapore},
  year      = {2018},
  pages     = {189--197}
}

@inproceedings{vishwarupe2016webbehavior,
  author    = {Saniya Zahoor and Mangesh Bedekar and Vinod Mane and Varad Vishwarupe},
  title     = {Uniqueness in User Behavior While Using the Web},
  booktitle = {Proceedings of the International Congress on Information and Communication Technology},
  publisher = {Springer Singapore},
  address   = {Singapore},
  year      = {2016},
  pages     = {221--228}
}

@inproceedings{bansal2021whole,
  author    = {Gagan Bansal and Tongshuang Wu and Joyce Zhou},
  title     = {Does the Whole Exceed its Parts? The Effect of {AI} Explanations on Complementary Team Performance},
  booktitle = {Proceedings of the 2021 {CHI} Conference on Human Factors in Computing Systems},
  year      = {2021},
  publisher = {Association for Computing Machinery},
  address   = {New York, NY, USA},
  doi       = {10.1145/3411764.3445088},
  articleno = {592},
  numpages  = {16}
}

@article{oliver1980cognitive,
  author  = {Richard L. Oliver},
  title   = {A Cognitive Model of the Antecedents and Consequences of Satisfaction Decisions},
  journal = {Journal of Marketing Research},
  volume  = {17},
  number  = {4},
  year    = {1980},
  pages   = {460--469}
}

@article{bhattacherjee2001understanding,
  author  = {Anol Bhattacherjee},
  title   = {Understanding Information Systems Continuance: An Expectation-Confirmation Model},
  journal = {MIS Quarterly},
  volume  = {25},
  number  = {3},
  year    = {2001},
  pages   = {351--370},
  doi     = {10.2307/3250921}
}

@inproceedings{passi2019problem,
  author    = {Samir Passi and Solon Barocas},
  title     = {Problem Formulation and Fairness},
  booktitle = {Proceedings of the Conference on Fairness, Accountability, and Transparency},
  year      = {2019},
  publisher = {Association for Computing Machinery},
  address   = {New York, NY, USA},
  doi       = {10.1145/3287560.3287567},
  pages     = {39--48}
}

@inproceedings{madaio2020codesigning,
  author    = {Michael A. Madaio and Luke Stark and Jennifer Wortman Vaughan and Hanna Wallach},
  title     = {Co-designing Checklists to Understand Organizational Challenges and Opportunities Around Fairness in {AI}},
  booktitle = {Proceedings of the 2020 {CHI} Conference on Human Factors in Computing Systems},
  year      = {2020},
  publisher = {Association for Computing Machinery},
  address   = {New York, NY, USA},
  doi       = {10.1145/3313831.3376445},
  articleno = {693},
  numpages  = {14}
}

@inproceedings{odriscoll2025socialnorms,
  author    = {Aoife O'Driscoll and Alan F. Blackwell},
  title     = {Social Norms, Social {AI}: Investigating the Effects of {AI} (Im)politeness and Gender on User Perception},
  booktitle = {Proceedings of BCS Human-Computer Interaction Conference 2025},
  year      = {2025},
  publisher = {BCS Learning \& Development Ltd.},
  doi       = {10.14236/ewic/BCSHCI2025.66},
  url       = {https://doi.org/10.14236/ewic/BCSHCI2025.66}
}

@article{vishwarupe2022xai,
author = {Vishwarupe, Varad and Joshi, Prachi M. and Mathias, Nicole and Maheshwari, Shrey and Mhaisalkar, Shweta and Pawar, Vishal},
title = {Explainable AI and Interpretable Machine Learning: A Case Study in Perspective},
journal = {Procedia Computer Science},
volume = {204},
pages = {869--876},
year = {2022},
doi = {10.1016/j.procs.2022.08.105}
}

@article{vishwarupe2022hcai,
author = {Vishwarupe, Varad and Maheshwari, Shrey and Deshmukh, Aseem and Mhaisalkar, Shweta and Joshi, Prachi M. and Mathias, Nicole},
title = {Bringing Humans at the Epicenter of Artificial Intelligence: A Confluence of AI, HCI and Human Centered Computing},
journal = {Procedia Computer Science},
volume = {204},
pages = {914--921},
year = {2022},
doi = {10.1016/j.procs.2022.08.111}
}

@incollection{vishwarupe2023industry4,
author = {Vishwarupe, Varad and Joshi, Prachi and Maheshwari, Shrey and Kuklani, Priyanka and Shingote, Prathamesh and Pande, Milind and Pawar, Vishal and Deshmukh, Aseem},
title = {Exploring Human Computer Interaction in Industry 4.0},
booktitle = {AI, IoT, Big Data and Cloud Computing for Industry 4.0},
publisher = {Springer},
year = {2023},
pages = {21--38},
doi = {10.1007/978-3-031-29713-7_2}
}
\appendix


\section{ APPENDIX A: Interview Guide}
Interviews were semi-structured and adaptive. Questions were used flexibly and adapted as analysis progressed, consistent with constructivist grounded theory. The examples below illustrate the types of prompts used; not all questions were answered in every interview depending on the comfort level of participants and time constraints.

\noindent\textbf{Background and role}
\begin{itemize}
\item Can you describe your role and how you work with AI or ML systems in your current position?
\item What kinds of decisions are you typically responsible for?
\end{itemize}

\noindent\textbf{Experiences with LLMs}
\begin{itemize}
\item Can you walk me through a recent situation where an LLM was considered or used in your work?
\item What problem was the LLM expected to help with?
\end{itemize}

\noindent\textbf{Decision-making and hesitation}
\begin{itemize}
\item Were there any moments where you or your team hesitated to use an LLM? What prompted that hesitation?
\item How were concerns or disagreements discussed within the team?
\end{itemize}

\noindent\textbf{Responsibility and oversight}
\begin{itemize}
\item Who was ultimately responsible for decisions or outputs involving the LLM?
\item How were outputs reviewed, checked, or justified?
\end{itemize}

\noindent\textbf{Boundaries and constraints}
\begin{itemize}
\item Were there limits placed on what the LLM could or could not do?
\item How were these boundaries decided?
\end{itemize}

\noindent\textbf{Reflection}
\begin{itemize}
\item Looking back, what made the use of the LLM feel acceptable or unacceptable in this case?
\item If you were to redesign this workflow, what would you change?
\end{itemize}

\section{APPENDIX B: CGT Coding Examples}
To support transparency in our constructivist grounded theory analysis, this appendix provides illustrative examples of how analytic categories were developed from interview data. Consistent with CGT principles, coding proceeded iteratively and concurrently with data collection, moving from initial open codes to more focused analytic categories through constant comparison and memo writing.

Initial open codes captured how participants reasoned about LLM use in concrete situations, including concerns about control, responsibility, justification, and organisational acceptability (e.g., P1 --- ``needing to stand behind decisions,'' P6 --- ``discomfort with shared agency,'' P18 --- ``keeping humans visibly in the loop,'' P30 --- ``difficulty explaining outcomes''). Through focused coding in NVivo, these were refined into higher-level analytic categories related to role positioning, accountability ownership, and oversight practices.

The table below provides illustrative examples of this analytic progression. These examples are not exhaustive, but demonstrate how participants' accounts informed the development of the grounded theory reported in the paper.

\begin{table}[t]
\centering
\caption{Coding examples (illustrative).}
\label{tab:coding_examples}
\begin{tabular}{p{0.20\linewidth}p{0.33\linewidth}p{0.22\linewidth}p{0.22\linewidth}}
\toprule
\textbf{Participant} & \textbf{Example excerpt} & \textbf{Initial open code} & \textbf{Focused category} \\
\midrule
P6 & ``I’m accountable, so I need to be able to stand behind the decision.'' & Owning final decision & Human-owned accountability \\
P9 & ``Once it feels like it’s collaborating, it’s unclear who’s responsible.'' & Ambiguous responsibility & Teammate role tension \\
P22 & ``We limited what it could do so reviews stayed human.'' & Constraining system scope & Tool-like role enforcement \\
P30 & ``It was hard to justify why the model did that.'' & Difficulty explaining output & Oversight breakdown \\
\bottomrule
\end{tabular}
\end{table}

These focused categories informed the analytic dimensions presented in Table~\ref{tab:rubric} and the role-positioning process illustrated in Fig.~\ref{fig:rolepositioning}. The paper is intended to stand independently of both appendices; the materials here are provided to clarify analytic practice rather than to enumerate all codes generated during analysis.

\end{document}